\documentclass[12pt]{article}
\usepackage[utf8]{inputenc}
\usepackage[english]{babel}
\usepackage{amsmath, amssymb}
\usepackage{booktabs}
\usepackage{graphicx}
\usepackage{hyperref}
\usepackage{float}
\usepackage{rotating}
\usepackage{natbib}

\title{\textbf{Poisson Regression under Multivariate Sample Selection}}

\author{
Kirill O. Morozov\\
HSE University\\
\texttt{kirmorozov6@gmail.com}
}

\newcommand{\E}{\mathbb{E}}
\renewcommand{\P}{\mathbb{P}}
\newcommand{\Var}{\operatorname{Var}}

\begin{document}
\maketitle

\begin{abstract}

This paper develops a Poisson regression model with multivariate sample selection, in which the outcome is observed only when several potentially correlated selection conditions are satisfied. To the best of our knowledge, this is the first Poisson sample selection model that allows for an arbitrary number of selection equations. We derive the conditional mean of the observed outcome under joint normality of the outcome and selection errors and obtain a multivariate selection-correction term.

We prove identification of the model parameters and show that, under suitable support and rank conditions, the outcome parameters can be identified without an exclusion restriction. We also propose two-step estimation procedures based on nonlinear least squares and Poisson pseudo-maximum likelihood. To the best of our knowledge, this paper is the first to apply PPML to a Poisson regression model with sample selection. The consistency of both estimators is established, and a robust two-step sandwich covariance matrix is proposed to account for the estimation error from the first-step selection model.

In addition, factorial moments are used to recover the variance of the latent outcome error and the correlations between the outcome and selection errors. Monte Carlo simulations show that ignoring sample selection leads to persistent bias when the outcome and selection errors are correlated, while the proposed PPML estimator substantially reduces this bias and is more stable than nonlinear least squares, especially under moderate and strong selection dependence.

\end{abstract}

\section{Introduction}
Count outcomes frequently arise in economics, particularly in fields such as health economics, labour economics, economics of crime, economics of trade among many others \citep{winkelmann2008econometric}. Since count variables are non-negative integers, conventional linear regression models are often inappropriate for their analysis. As a result, econometric models specifically designed for count data have received considerable attention in the literature \citep{winkelmancountd,hausman,haslett2022modelling}.

The Poisson regression model remains the classic specification for the analysis of count data \citep{cameron2013regression}. However, this model assumes that the conditional variance is equal to the conditional mean, which is often inconsistent with empirical applications. Consequently, more flexible specifications such as the negative binomial model have become widely used in applied research \citep{hausman,hilbe}.

In many empirical applications, the outcome variable is observed only for a non-randomly selected subset of the population. Such sample selection mechanisms arise when the probability of observing an outcome depends on observed or unobserved characteristics that also affect the outcome of interest. For example, the number of trips taken by household members may only be observed for households satisfying specific survey participation conditions, such as vehicle ownership requirements.
Ignoring this non-random selection process may therefore lead to biased and inconsistent parameter estimates \citep{terza1998}.

In the patent-production framework  \citep{hausman}, observed patent counts are modeled as the outcome of a single stochastic count-generating process. However, the actual innovation mechanism is substantially more complex. Before a patent count becomes observable, a firm must pass through several latent stages, including engagement in R\&D activity, generation of patentable innovations, submission of patent applications, and successful patent approval. These stages are driven by different economic factors and may involve distinct dependence structures. Standard count-data sample selection models typically collapse these mechanisms into a single latent participation equation, implicitly assuming that selection occurs through one binary process only \citep{greene1994,winkelmann1998}. Such a specification may be overly restrictive in empirical environments where observation of count outcomes is generated by multiple sequential or interdependent selection mechanisms. This motivates the development of generalized count-data selection frameworks with multiple latent selection equations.

The classical sample selection model for continuous outcomes was introduced by
\citet{heckman1979}. Heckman proposed a procedure that accounts for non-random
sample selection and allows consistent estimation of model parameters. However,
the original framework considered only a single outcome equation and a binary
selection mechanism. Since then, the sample selection framework has been extended
to more general settings with multiple outcomes, multiple selection equations,
and more complex selection mechanisms
\citep{Poirier, DMF,
Bourguignon,Novi,Tauchmann,DeLuca,Li,Ogundimu,Kim,Kossova2018,Kupriianova2020,Rezaee,Kossova2022}.
Nevertheless, these approaches were primarily developed for continuous outcomes
and cannot be directly applied to count-data settings.

Several approaches have been proposed to incorporate sample selection mechanisms into count-data models. \citet{greene1994} extended Poisson and negative binomial specifications to account for non-random sample selection, excess zeros, and overdispersion in count outcomes. \citet{terza1998} developed count-data models with endogenous switching, sample selection, and endogenous treatment effects, and proposed both fully parametric and partially parametric estimation methods. \citet{miranda2006} further extended maximum likelihood estimation to endogenous switching and sample selection models for count outcomes, which made these models more accessible for applied empirical work. More recently, \citet{wysz2018} proposed a flexible semiparametric copula-based framework for count-data sample selection models, allowing for non-normal dependence between the selection and outcome equations, nonlinear covariate effects, and alternative count distributions. These contributions substantially advanced the modeling of count data under non-random selection. 

However, they remain based on the standard two-equation structure, in which a single binary selection equation determines whether the count outcome is observed. Thus, the existing literature does not provide a general framework that allows for an arbitrary number of selection equations. This limitation is important when sample inclusion is generated by several sequential or simultaneous selection mechanisms. The present paper addresses this gap by developing a generalized count-data sample selection model with an arbitrary number of selection equations.

\section{Count data model with multivariate sample selection}

\subsection{Model}

Throughout the rest of the paper, we assume that the errors are independent across observations. The index j denotes the selection equation.

Consider binary selection variables $z_{i,j}$ such that:
\begin{equation*}
z_{i,j} = \begin{cases}
   1 ,~w_{i,j}\gamma_j+u_{i,j}\geq0\\
   0 ,~\text{otherwise}
 \end{cases}=
 \begin{cases}
   1,~-w_{i,j}\gamma_j\leq u_{i,j}\\
   0 ,~\text{otherwise},
 \end{cases}
\end{equation*}
where $w_{i,j}$  is a row vector of regressors, $\gamma_j$ is a column vector of coefficients and $u_j$ is a random error. 

Denote by $y_i^*$ and $y_i$ a latent dependent (target) variable and its censored value, respectively. Specifically, following Heckman (1979), Kossova \& Potanin (2018). Suppose that $y_i$ is observable only for particular values of this binary variable

\begin{equation*}
\begin{aligned}
y_i
&=
\begin{cases}
y_i^*, & \text{if } z_{i,1}=1,\ldots,z_{i,m}=1,\\
\text{unobserved}, & \text{otherwise},
\end{cases}
\\[0.5em]
&=
\begin{cases}
y_i^*, & \text{if } -\eta_{i,1}\leq u_{i,1},\ldots,-\eta_{i,m}\leq u_{i,m},\\
\text{unobserved}, & \text{otherwise},
\end{cases}
\end{aligned}
\end{equation*}

We will write $z_{i,1}=1,\ldots,z_{i,m}=1$ as $Z_i = 1$ for brevity.

in contrast to previous studies on multivariate sample selection we assume that $y^*_i|~x_i, \varepsilon_i $ is taken from a discrete distribution with $\E[y^*|x_i,\varepsilon]=\text{exp}\{x_i\beta+\varepsilon_i\}$. Throughot the article $x_i$ is a row-vector of regressors, $\beta$ - is a column-vector m$\times$1.

For example, suppose that $y_i$ is the number of cigarettes smoked per day. We observe this variable only if the respondent smokes and answers the question about the number of cigarettes smoked. Thus, $z_{i,1}$ indicates whether the respondent smokes, while $z_{i,2}$ indicates whether the respondent answered the corresponding question.

Let
\(
u_i=(u_{i,1},\ldots,u_{i,m})~\text{and }
\eta_i= (w_{i, 1}\gamma_1, \ldots, w_{i, m}\gamma_m)
\). Then the multiple selection event can be written as
\(
u_i \geq -\eta_i,
\)
where the inequality is understood element by element.

Following \citep{Das2003,Kossova2018} suppose that ($\varepsilon_i$, \ldots, $u_{i,m})'$ are independent of ($x_i$ $w_i$). Also $\E[y_i^*|x_i,w_i,\varepsilon_i,u] = \E[y^*_i|x_i,\varepsilon_i] = \text{exp}\{x_i\beta+\varepsilon_i\}$. Under this assumption, the variables and random errors entering the selection equations do not directly affect the outcome once \(\varepsilon_i\) is fixed. However, \(\varepsilon_i\) and $u_i$ may be correlated.
 Then by the law of iterated expectations:

\begin{equation}
\begin{aligned}
&\E\!\left[
y^*_i
\mid
x_i,w_i,Z_i=1
\right] 
\\&=
\E\!\left[
\E\!\left[
y^*_i
\mid
x_i,w_i,Z_i=1,
\varepsilon_i,u_i
\right]
\mid
x_i,w_i,Z_i=1
\right]
\\
&=
\E\!\left[
\E\!\left[
y^*_i
\mid
x_i,w_i,
\varepsilon_i,u_i
\right] 
\mid
x_i,w_i,Z_i=1
\right]
\\
&=
\E\!\left[
e^{x_i\beta+\varepsilon_i}
\mid
x_i,w_i,Z_i=1
\right]
\\
&=
\E\!\left[
e^{x_i\beta+\varepsilon_i}
\mid
x_i,w_i,u_i\geq -\eta_i
\right] =
e^{x_i\beta}
\E\!\left[
e^{\varepsilon_i}
\mid x_i, w_i,
u_i\geq -\eta_i
\right].
\end{aligned}
\end{equation}

 $Z_i=1$ is fully determined by $w_i$ and $u_i$, since $z_{i,j}=1$ if and only if $u_{i,j}\geq -w_{ij}\gamma_j$. Therefore, conditional on $w_i$ and $u_i$, this event gives no additional information.

Notice that $\E\!\left[
e^{\varepsilon_i}
\mid
u_i\geq -\eta_i
\right]$ depends on $-\eta_i$ that depends on $w_i$, so the expectation is not a constant but the function of $\eta_i$. We consider two estimators which address  this issue.

Suppose that the joint distribution of \(\varepsilon_i\) and \(u_i\) is multivariate normal:
\[
\begin{pmatrix}
\varepsilon_i \\
u_{i,1} \\
u_{i,2} \\
\vdots \\
u_{i,m}
\end{pmatrix}
\sim
\mathcal{N}
\left[
\begin{pmatrix}
0 \\
0 \\
0 \\
\vdots \\
0
\end{pmatrix},
\begin{pmatrix}
\sigma^2 & \delta_1 & \delta_2 & \cdots & \delta_m \\
\delta_1 & 1 & \rho_{1,2} & \cdots & \rho_{1,m} \\
\delta_2 & \rho_{2,1} & 1 & \cdots & \rho_{2,m} \\
\vdots & \vdots & \vdots & \ddots & \vdots \\
\delta_m & \rho_{m,1} & \rho_{m,2} & \cdots & 1
\end{pmatrix}
\right].
\]
Here \(\delta=(\delta_1,\ldots,\delta_m)'\) is the covariance vector between the outcome error and the selection errors, that is \(\delta_j=\operatorname{Cov}(\varepsilon_i,u_{i,j})\). Since \(\operatorname{Var}(u_{i,j})=1\), the corresponding correlation is \(\rho_j=\dfrac{\delta_j}{\sigma}\). In the conditional-mean representation below we use \(\delta\), while the separate of \(\sigma\) and \(\rho\) is discussed later. In an ordered probit model, the scale of the latent variable is not identified from the observed ordered outcomes. Therefore, the variance of the latent error is normalized to one, $\operatorname{Var}(u_{i,j})=1$, which fixes the scale of the model; see \citet{Hansen2022}.

For  calculation $\E\!\left[
e^{\varepsilon_i}
\mid
u_i\geq -\eta_i
\right]$ we use moment-generating function of multivariate normal truncated distribution \citep{wilhelm2021} which is provided in Appendix~\ref{app:mgf}. 

By definition m.g.f of random column-vector $\xi$ is $\E[\text{exp}\{t'\xi\}]$ where t is same space column-vector so for truncated distribution it takes the form $\E[\text{exp}\{t'\xi\}|a\leq\xi\leq b]$, where $a$ and $b$ are vectors of lower and upper truncation points, respectively. In our case there is only bottom truncation, so $b$ goes to $+\infty$, $\varepsilon$ is not truncated, and all other variables are truncated by $-\eta_i$. So when $t=(1 ~0~\cdots ~0)$ and $a = (-\infty ~-\eta_{i,1} ~\ldots ~ -\eta_{i,m} )$ then we may get the conditional expected value, but for the later decomposition, it is useful to write this expression not only for \(r=1\), but for any positive integer \(r\). Taking \(t_r=(r,0,\ldots,0)'\), the moment-generating function of the truncated multivariate normal distribution gives in formula \ref{eq:truncated_mgf}

\begin{equation} \begin{aligned} &\E\left[ \exp\{r\varepsilon_i\} \mid u_{i1}\geq -\eta_{i,1},\ldots,u_{i,m}\geq -\eta_{i,m} \right] \\ &= \frac{\exp\left\{\frac{r^2\sigma^2}{2}\right\}} {\alpha(2\pi)^{\frac{m+1}{2}}|\Sigma_{(m+1)\times(m+1)}|^{\frac{1}{2}}}\times \\ & \int_{-\infty}^{+\infty} \ldots \int_{-\eta_{i,m}-r\delta_m}^{+\infty} \exp\left\{ -\frac{1}{2}x'\Sigma_{(m+1)\times(m+1)}^{-1}x \right\} dx_{u_m}\cdots dx_{\varepsilon} \\ &= \frac{ \exp\left\{\frac{r^2\sigma^2}{2}\right\} \cdot \P\{-(\eta_{i,m}+r\delta_m)<u_{i,m},\ldots, -(\eta_{i,1}+r\delta_1)<u_{i,1}\} } { \P\{-\eta_{i,m}<u_{i,m},\ldots,-\eta_{i,1}<u_{i,1}\} } \\ &= \exp\left\{\frac{r^2\sigma^2}{2}\right\} \cdot \frac{ \Phi_m(\eta_{i,m}+r\delta_m,\ldots,\eta_{i,1}+r\delta_1) }{ \Phi_m(\eta_{i,m},\ldots,\eta_{i,1}) }. \end{aligned} \label{eq:gencor} \end{equation}
where $\alpha$ = $\P\{-\eta_{i,m}<u_{i,m}, \ldots, -\eta_{i,1}<u_{i,1}\}$, $\Phi_m(\cdot)$ is a cumulative distribution function of the $m$-dimensional normal distribution with mean vector $\mathbf{0}$ and covariance matrix $\Sigma_{m\times m}$.

This expression also includes the conditional mean correction as a special case. Indeed, setting \(r=1\) gives the \(\E[\text{exp}\{\varepsilon_i\}\mid x_i,w_i,z_{i,1}=1,\ldots,z_{i,m}=1]\). For higher values of \(r\), the same formula will be used to construct factorial moment identities that will be used for decomposition between correlation of selection and outcome equations and variance of outcome equation.

Finally, using equation~\eqref{eq:gencor}, we obtain the following result:
\begin{equation}
    \E[y^*_i \mid x_i,w_i, z_{i,1}=1,\ldots,z_{i,m}=1]
    =
    e^{x_i\beta+\frac{\sigma^2}{2}}
    \cdot
    \frac{
    \Phi_m(w_{i,m}\gamma_m+\delta_m,\ldots,w_{i,1}\gamma_1+\delta_1)
    }{
    \Phi_m(w_{i,m}\gamma_m,\ldots,w_{i,1}\gamma_1)
    }.
\end{equation}

For brevity, denote the selection-correction term as
\[
\lambda(w_i\gamma,\delta;R)
=
\frac{
\Phi_m(w_{i,m}\gamma_m+\delta_m,\ldots,w_{i,1}\gamma_1+\delta_1;R)
}{
\Phi_m(w_{i,m}\gamma_m,\ldots,w_{i,1}\gamma_1;R)
},
\]
 where \(w_i\gamma=(w_{i,1}\gamma_1,\ldots,w_{i,m}\gamma_m)\), \(\delta=(\delta_1,\ldots,\delta_m)\) are regressors from selection equations and the covariance vector between the outcome error, and \(R\) is the correlation matrix of the selection errors.
The obtained results generalize the model of \citet{terza1998}. When there is only one selection equation, the conditional expectation is the same as in the original model.

\subsection{Identification of the generalized count data sample selection model}
~\label{sec:identification}

Under the following sufficient conditions, the outcome parameters are identified without an exclusion restriction:

\begin{itemize}
\item[\textbf{A1.}] The multivariate probit selection subsystem is identified.
\item[\textbf{A2.}] The disturbances are jointly normally distributed as specified above, with $R$ positive definite and $\sigma^2>0$.
\item[\textbf{A3.}] The support of the continuous regressor vector $Z$ contains a nonempty open subset of $\mathbb{R}^k$.
\item[\textbf{A4.}] The matrix of coefficients determining the selection indices has full row rank, $\operatorname{rank}(A)=m$.
\end{itemize}

\noindent
Under A1--A4, if two parameter pairs $(\beta,\delta)$ and
$(\widetilde{\beta},\widetilde{\delta})$ generate the same conditional mean for all admissible values of the regressors, then

$$
\widetilde{\beta}=\beta,
\qquad
\widetilde{\delta}=\delta.
$$

Hence, the outcome parameters are identified without an exclusion restriction.

The identification result follows from the nonlinear form of the multivariate normal selection correction and the support and rank conditions stated above. A formal proof is provided in Appendix ~\ref{app:identification}.

\subsection{Decomposition of structural parameters}

The conditional-mean representation identifies \((\beta,\gamma,R,\delta)\), but it does not separately recover the variance \(\sigma^2\) of the outcome error and the correlations \(\rho_j\). Getting $\rho$ is crucial for economic interpretation as power of connection between unobserved factors that affects both the count equation and selection equatinos. To recover these parameters, we use the normalized second factorial moment of $y_i$ - \(\widehat{A}_2\). Its full derivation is presented in Appendix \ref{app:decomp}, while here we report only the expressions used for calculation.

The estimation of The normalized second factorial moment is defined as
\[
\widehat{A}_2
=
\dfrac{1}{n_s}
\sum_{i=1}^{n}
\mathbb{I}\{Z_i=1\}
\dfrac{(y_i)_2}{\widehat{\mu}_i^2}
\dfrac{\widehat{\lambda}_{1,i}^2}
{\widehat{\lambda}_{2,i}},
\]
where \(n\) is the total number of observations, \(n_s=\sum_{i=1}^{n}\mathbb{I}\{Z_i=1\}\) is the number of selected observations, where $\mathbb{I}$ is indicator function. The second falling factorial is \((y_i)_2=y_i(y_i-1)\), while \(\widehat{\mu}_i\) is the fitted conditional mean. The selection-correction term of order \(r\) is defined as
\[
\widehat{\lambda}_{r,i}
=
\dfrac{
\Phi_m\left(
w_i\widehat{\gamma}
+
r\widehat{\delta};
\widehat{R}
\right)
}{
\Phi_m\left(
w_i\widehat{\gamma};
\widehat{R}
\right)
}.
\]

For the Poisson model, the normalized second factorial moment satisfies \(\widehat{A}_2=\exp\{\widehat{\sigma}^2\}\). Therefore, the variance of the latent outcome error is recovered as \(\widehat{\sigma}^2=\ln\widehat{A}_2\), while its standard deviation is \(\widehat{\sigma}=\sqrt{\ln\widehat{A}_2}\).

Since \(\delta_j=\sigma\rho_j\), the correlations between the outcome error and the selection errors are estimated  as \(\widehat{\rho}_j=\dfrac{\widehat{\delta}_j}{\widehat{\sigma}}\).

If the outcome equation contains a constant, the structural constant is recovered as \(\widehat{\beta}^{\,str}_0=\widehat{\beta}_0-\dfrac{\widehat{\sigma}^2}{2}\).

\section{Estimation of count data multiple sample selection generalization}

\subsection{Nonlinear least squares}
The estimation can be implemented in two steps the same way is shown at \citet{terza1998}. At the first step, we estimate
the selection subsystem. Since the selection equations have the form
$z_{i,j}=\mathbb{I}\{w_{i,j}\gamma_j+u_{i,j}\geq 0\}$, $j=1,\ldots,m$ and we suggest that errors are normally distributed, this part of the
model is a standard multivariate probit model. Therefore, the parameters
$\gamma$ and the correlation matrix $R$ can be estimated by maximum
likelihood using all observations.

At the second step, we estimate the parameters of the outcome equation.
Then $\beta$ and $\delta$
can be estimated by minimizing the following nonlinear least squares
criterion:
\[
S(\hat{\beta},\hat{\rho})
=
\arg\min_{\beta,\rho}
\frac{1}{n_s}
\sum_{i=0}^n
\mathbb{I}\{Z_i-1\}\left[
y_i-\exp\{x_i\beta\}\lambda(w_i\hat{\gamma},\delta)
\right]^2,
\]

Since the second step minimizes the sample average of the loss function and the first-step parameters are consistent, the proposed procedure can be treated as a two-step M-estimator in the sense of Section 12 of \citet{wooldridge}. Below, we show that the sufficient conditions for consistency are satisfied in the present model.

 Let $\theta=(\beta',\delta')'$ be the vector of second-step parameters. Denote the sample criterion by $S_n(\theta)$ and define the theoretical criterion as
$$
S(\theta)
=
\mathbb{E}\left[
\left(y_i-\exp\{x_i\beta\}\lambda(w_i{\gamma},\delta)\right)^2
\mid
z_{i,1}=\cdots=z_{i,m}=1
\right].
$$
 Under the regularity conditions stated by
Theorem 22.2 of \citet{Hansen2022} implies that the sample criterion uniformly converges in probability to the theoretical criterion on $\Theta$, that is 
$$
\sup_{\theta\in\Theta}
\left|S_n(\theta)-S(\theta)\right|
\overset{p}{\longrightarrow}
0.
$$

Therefore, asymptotically, minimization of the sample criterion is equivalent to minimization of the theoretical criterion.

By Theorem 2.7 of \citet{Hansen2022}, the theoretical mean squared error is minimized by the conditional expectation function. In the present model this function is $m_0(x_i,w_i)=\mathbb{E}[y_i\mid x_i,w_i,z_{i,1}=\cdots=z_{i,m}=1]=g(x_i,w_i;\theta_0)$. Hence, the theoretical criterion can attain its minimum only if $g(x_i,w_i;\theta)=m_0(x_i,w_i)$ almost surely on the selected sample. Since the model is identified, this equality implies $\theta=\theta_0$. Therefore, $S(\theta)$ has a unique minimum at the true parameter vector $\theta_0$.

Therefore, the proposed estimation procedure can be interpreted as a two-step M-estimator in the sense of \citet{wooldridge}. The first step provides consistent estimates of the selection-equation parameters, while the second step minimizes the nonlinear least squares criterion with generated first-step quantities. Under the standard regularity conditions for two-step M-estimators, $S_n(\theta)$ uniformly converges to $S(\theta)$. Since $S(\theta)$ is uniquely minimized at $\theta_0$, the second-step nonlinear least squares estimator is consistent:
$$
(\hat{\beta},\hat{\delta})
\overset{p}{\longrightarrow}
(\beta_0,\delta_0).
$$

\subsection{Poisson pseudo-maximum likelihood}

As an alternative to nonlinear least squares, the parameters of the outcome equation can be estimated by Poisson pseudo-maximum likelihood (PPML). \citet{santos} show that PPML can be consistently applied when the conditional mean is correctly specified, without requiring the dependent variable to follow a Poisson distribution. We adapt this approach to the present sample selection model. Thus, under standard regularity conditions, consistency of the second-step estimator requires correct specification of the conditional mean rather than the full conditional distribution.

Let \(\theta=(\beta',\delta')'\) denote the vector of second-step parameters, and let \(\eta=(\gamma,R)\) are parameters estimated at the first step. For selected observations, define the conditional mean as \(m_i(\theta,\eta)=\exp\{x_i\beta\}\lambda(w_i\gamma,\delta;R)\).

The population PPML objective function for the selected observations is

\[
Q(\theta,\eta_0)
=
\mathbb{E}\left[
m_i(\theta,\eta_0)
-
y_i\ln m_i(\theta,\eta_0)
\mid Z_i=1
\right].
\]

At the true parameter vector, the corresponding population moment condition
is equal to zero. By the law of iterated expectations,

\[
\begin{aligned}
&\mathbb{E}\left[
\left.
\dfrac{\partial \ln m_i(\theta_0,\eta_0)}{\partial\theta}
\left[
y_i-m_i(\theta_0,\eta_0)
\right]
\right|
Z_i=1
\right]
\\
&=
\mathbb{E}\left[
\left.
\dfrac{\partial \ln m_i(\theta_0,\eta_0)}{\partial\theta}
\underbrace{
\mathbb{E}\left[
y_i-m_i(\theta_0,\eta_0)
\mid x_i,w_i,Z_i=1
\right]
}_{=\,0}
\right|
Z_i=1
\right]
=0,
\end{aligned}
\]

because

\[
m_i(\theta_0,\eta_0)
=
\mathbb{E}[y_i\mid x_i,w_i,Z_i=1].
\]

The sample analogue of the population objective function is

\[
Q_n(\theta,\widehat{\eta})
=
\dfrac{1}{n_s}
\sum_{i=1}^{n}
\mathbb{I}\{Z_i=1\}
\left[
m_i(\theta,\widehat{\eta})
-
y_i\ln m_i(\theta,\widehat{\eta})
\right].
\]

The PPML estimator minimizes this sample objective function. The corresponding
first-order condition is

\[
\dfrac{1}{n_s}
\sum_{i=1}^{n}
\mathbb{I}\{Z_i=1\}
\dfrac{\partial \ln m_i(\theta,\widehat{\eta})}{\partial\theta}
\left[
y_i-m_i(\theta,\widehat{\eta})
\right]
\rightarrow
0.
\]

We have already shown at Paragraph 2.2 that the conditional mean is identified. In particular, under the identification of the multivariate probit selection subsystem, the exclusion restriction, and the full-rank condition for the outcome regressors, \(m_i(\theta,\eta_0)=m_i(\theta_0,\eta_0)\) almost surely implies \(\theta=\theta_0\).

It remains to show that the population PPML criterion is uniquely minimized at the true parameter vector. Define
\[
Q(\theta)
=
\mathbb{E}
\left[
m_i(\theta,\eta_0)
-
y_i\ln m_i(\theta,\eta_0)
\mid
Z_i=1
\right].
\]
Let \(m_{0i}=m_i(\theta_0,\eta_0)=\mathbb{E}[y_i\mid x_i,w_i,Z_i=1]\). Using the law of iterated expectations,
\[
Q(\theta)-Q(\theta_0)
=
\mathbb{E}
\left[
m_i(\theta,\eta_0)-m_{0i}
-
m_{0i}
\ln
\left(
\dfrac{m_i(\theta,\eta_0)}{m_{0i}}
\right)
\Bigm|
Z_i=1
\right].
\]

Define \(a_i(\theta)=\dfrac{m_i(\theta,\eta_0)}{m_{0i}}\). Since both conditional means are strictly positive, \(a_i(\theta)>0\), and therefore
\[
Q(\theta)-Q(\theta_0)
=
\mathbb{E}
\left[
m_{0i}
\left\{
a_i(\theta)-1-\ln a_i(\theta)
\right\}
\mid
Z_i=1
\right].
\]
For every \(a>0\), we have \(a-1-\ln a\geq 0\), with equality if and only if \(a=1\). Hence, \(Q(\theta)\geq Q(\theta_0)\), and equality is possible only if \(m_i(\theta,\eta_0)=m_{0i}\) almost surely. Since the conditional mean is identified, this equality implies \(\theta=\theta_0\). Therefore, the population PPML criterion has a unique global minimum at the true parameter vector.

The PPML estimator is an M-estimator because it minimizes a sample average of the pseudo-likelihood loss function. Since the first-step estimator is consistent and, under the standard regularity conditions for two-step M-estimators
\[
\sup_{\theta\in\Theta}
\left|
Q_n(\theta,\widehat{\eta})
-
Q(\theta,\eta_0)
\right|
\overset{p}{\longrightarrow}
0,
\]
the consistency results for two-step M-estimators in Section~12 of \citet{wooldridge} apply. Since \(Q(\theta)\) is uniquely minimized at \(\theta_0\), the second-step PPML estimator is consistent:
\[
(\widehat{\beta},\widehat{\delta})
\overset{p}{\longrightarrow}
(\beta_0,\delta_0).
\]

\subsection{Estimator of the asymptotic covariance matrix }

Since $(\lambda(w_i\widehat{\gamma},\rho;\widehat{R}))$ is constructed using first-step estimates, the usual nonlinear least squares covariance matrix is not sufficient. It treats $(\widehat{\gamma})$ and $(\widehat{R})$ as fixed, although their estimation error also affects $(\widehat{\beta}) $and $(\widehat{\rho})$. Therefore, to address for the generated regressors we proposed the robust two-step sandwich covariance matrix.

Then the covariance matrix is estimated as
\[
\widehat{\mathrm{Var}}(\widehat{\theta})
=
\frac{1}{n}
\widehat{H}^{-1}
\widehat{\Omega}
(\widehat{H}^{-1})',
\]
where
\[
\widehat{H}
=
\frac{1}{n}
\sum_{i=1}^{n}
\frac{\partial \psi_i(\widehat{\theta})}{\partial \theta'},
\qquad
\widehat{\Omega}
=
\frac{1}{n}
\sum_{i=1}^{n}
\psi_i(\widehat{\theta})\psi_i(\widehat{\theta})'.
\]
The vector $\psi_i(\theta)$ contains the first-order conditions from the first-step likelihood and the second-step nonlinear least squares criterion. The detailed derivation of this covariance matrix is given in Appendix D. For inference on the outcome equation, we use the block of $\widehat{\mathrm{Var}}(\widehat{\theta})$ corresponding to $(\widehat{\beta}',\widehat{\rho}')'$.

The covariance matrix can also be used to test whether the selection correction is necessary by testing the joint null hypothesis \(H_0:\rho_1=\cdots=\rho_m=0\), for example, using a standard Wald test.

\section{Simulation example}

Unlike a conventional Monte Carlo design based on a single fixed parameterization, we allow several nuisance features of the data-generating process to vary across replications. This choice is deliberate. The purpose of the simulation is not to evaluate the estimators at one particular calibration, but to examine whether their relative performance is robust across a broader class of admissible data-generating processes. Similar simulation strategies have been used to evaluate estimator performance across multiple or randomly parameterized DGPs rather than at a single point in the parameter space \citep{li2024local,reuvers2024sparse}. Accordingly, the regression coefficients and the correlation structure of the regressors are allowed to vary across replications, while the main parameters of interest for the comparison, including the sample size and the strength of selection dependence, are controlled across simulation scenarios.

Thus, the reported Monte Carlo measures should be interpreted as average finite-sample performance over the specified distribution of DGPs rather than as performance conditional on a single fixed parameter vector.

\subsection{Monte Carlo design}

We use a Monte Carlo simulation with a Poisson outcome and two selection equations. In each replication, five regressors are generated as \(x_i=(x_{i,1},\ldots,x_{i,5})'\sim N(0,R_x)\). The correlation matrix \(R_x\) is generated separately in each replication using the method of \citet{joe2006}. 

The latent outcome is generated from
\[
y_i^*\mid x_i,\varepsilon_i
\sim
\operatorname{Poisson}
\left(
\exp\left\{
\beta_0^{str}
+\beta_1x_{i,1}
+\beta_2x_{i,2}
+\beta_3x_{i,3}
+\varepsilon_i
\right\}
\right).
\]
The coefficients \(\beta_0^{str},\beta_1,\beta_2,\beta_3\) are independently drawn from the uniform distribution \(U[-1,1]\) in each replication.

The selection equations are
\[
z_{i,1}
=
\mathbb{I}
\left\{
\gamma_{1,0}
+\gamma_{1,1}x_{i,1}
+\gamma_{1,4}x_{i,4}
+u_{i,1}
\geq 0
\right\},
\]
and
\[
z_{i,2}
=
\mathbb{I}
\left\{
\gamma_{2,0}
+\gamma_{2,2}x_{i,2}
+\gamma_{2,5}x_{i,5}
+u_{i,2}
\geq 0
\right\}.
\]

The outcome is observed only when \(z_{i,1}=z_{i,2}=1\). The variables \(x_{i,4}\) and \(x_{i,5}\) enter only the selection equations and are used as exclusion restrictions. The intercepts in the selection equations are drawn from \(U[0,1]\), while the remaining nonzero coefficients are drawn from \(U[-1,1]\).

The errors are jointly normally distributed:
\[
\begin{pmatrix}
\varepsilon_i\\
u_{i,1}\\
u_{i,2}
\end{pmatrix}
\sim
\mathcal{N}\left[
\begin{pmatrix}
0\\
0\\
0
\end{pmatrix},
\begin{pmatrix}
\sigma^2 & \sigma r & \sigma \rho\\
\sigma \rho & 1 & 0.6\\
\sigma \rho & 0.6 & 1
\end{pmatrix}
\right].
\]
We set \(\sigma=0.75\) and \(\operatorname{Corr}(u_{i,1},u_{i,2})=0.6\). The correlations between the outcome error and both selection errors are equal:
\[
\operatorname{Corr}(\varepsilon_i,u_{i,1})
=
\operatorname{Corr}(\varepsilon_i,u_{i,2})
=
\rho,
\qquad
\rho\in\{0,0.4,0.8\}.
\]
Therefore, \(\delta_1=\delta_2=\sigma \rho\).

We compare six estimators. The first two are naive Poisson and negative binomial regressions, which ignore sample selection. The next two use one selection equation for the combined indicator \(s_i=\mathbb{I}\{z_{i,1}=z_{i,2}=1\}\) and estimate the second step by NLS or PPML. The last two estimators use two separate selection equations and estimate the second step by NLS or PPML.

Each scenario is repeated \(B=100\) times. Let \(\widehat{\theta}_b\) and \(\theta_b\) denote the estimated and true values of a parameter in replication \(b\). Bias is calculated as
\[
\operatorname{Bias}(\widehat{\theta})
=
\dfrac{1}{B}
\sum_{b=1}^{B}
\left(
\widehat{\theta}_b-\theta_b
\right).
\]
The root mean squared error is
\[
\operatorname{RMSE}(\widehat{\theta})
=
\sqrt{
\dfrac{1}{B}
\sum_{b=1}^{B}
\left(
\widehat{\theta}_b-\theta_b
\right)^2
}.
\]
The root median squared error is
\[
\operatorname{RMedSE}(\widehat{\theta})
=
\sqrt{
\operatorname{median}_{b=1,\ldots,B}
\left\{
\left(
\widehat{\theta}_b-\theta_b
\right)^2
\right\}
}.
\]
The win rate is calculated as the share of common successful replications
in which an estimator has the smallest absolute error for a given parameter.
Formally,

\[
\operatorname{WinRate}_m
=
\frac{1}{B}
\sum_{b=1}^{B}
\frac{
\mathbb{I}\left\{
\left|\hat{\theta}_{m,b}-\theta_b\right|
=
\min_j
\left|\hat{\theta}_{j,b}-\theta_b\right|
\right\}
}{K_b},
\]

where \(K_b\) is the number of estimators
that attain the smallest absolute error in replication \(b\).
Thus, in case of ties, the win is divided equally among the tied estimators. 

\subsection{Monte Carlo results}

Figures 1--4 summarize the main Monte Carlo results. The complete numerical values of Bias, RMSE, RMedSE, and win rate are reported in Appendix \ref{app:mntc}.

Figure 1 shows that the importance of the selection correction depends directly on the correlation between the outcome and selection errors. When $\rho=0$, the naive Poisson and negative binomial estimators have almost zero bias. In this case, adding the selection terms does not correct any systematic error and only introduces additional estimation noise. Moreover, including the selection-correction term may introduce additional finite-sample bias. When $\rho$ increases to $0.4$ and $0.8$, the situation changes. the bias of the estimator of the intercept if naive models is approximately $0.15$ for $\rho=0.4$ and $0.28$ for $\rho=0.8$, and it does not decrease as $n$ grows. Thus, this bias is caused by sample selection and cannot be removed by increasing the sample size.

Both PPML selection estimators substantially reduce the bias caused by sample selection. For example, at \(\rho=0.8\), the intercept bias of the two-selection PPML estimator remains close to zero for all sample sizes, while the naive Poisson and negative binomial estimators remain strongly biased. A similar pattern is observed for the slope coefficients, although the bias is smaller than for the intercept. In contrast, the two-selection NLS estimator is less stable: its bias may increase or even change sign as \(n\) grows.

\begin{figure}[H]
    \centering
    \includegraphics[width=0.72\textwidth]{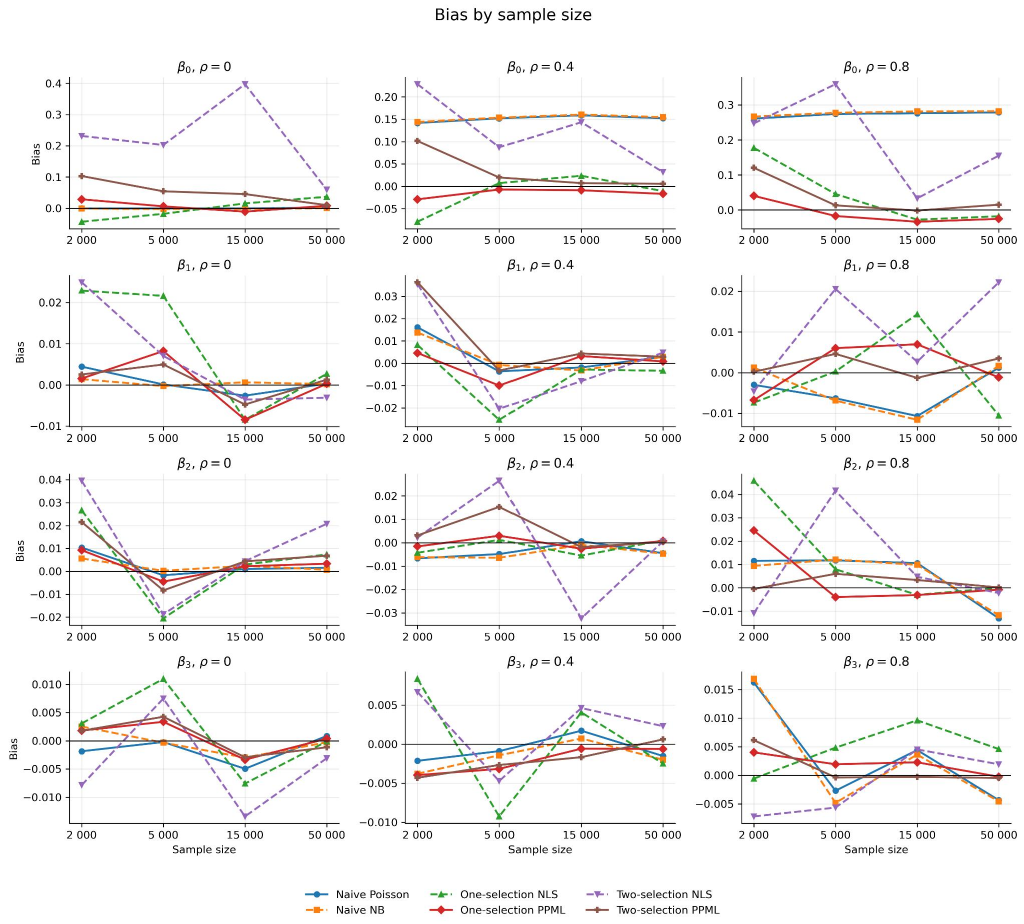}
    \caption{Bias of the Monte Carlo estimators by sample size and dependence level}
    \label{fig:mc_bias}
\end{figure}

Figure 2 gives a clearer comparison of the total estimation error. When $\rho=0$, the naive negative binomial estimator generally has the lowest RMSE. Therefore, estimating a selection correction when no selection dependence is present leads to an efficiency loss. For $\rho=0.4$, the two-selection PPML estimator improves rapidly as the sample size increases. For example, At $n=50{,}000$, it has a lower RMSE than the naive estimators for all four coefficients.

The advantage of the two-selection PPML estimator is strongest when $\rho=0.8$. For the intercept, its RMSE decreases from approximately $0.16$ at $n=5{,}000$ to $0.06$ at $n=50{,}000$. Over the same range, the RMSE of the naive estimators remains close to $0.28$. For the slope coefficients, the RMSE of the two-selection PPML estimator also has stable decreasing and becomes the lowest or nearly the lowest at $n=50{,}000$.

The NLS estimators are considerably less stable. For example, the RMSE of the two-selection NLS estimator for the intercept exceeds one in several scenarios and can increase even when the sample size grows. This instability is especially important for the structural intercept. It is recovered as
$\widehat{\beta}^{\,str}_0=\widehat{\beta}_0-\widehat{\sigma}^2/2$, so estimation error in $\widehat{\sigma}^2$ creates an additional source of error in the final intercept estimate. 

One possible explanation for the instability of NLS is the strong heteroskedasticity of the count outcome. Since the variance of the outcome increases with its conditional mean, observations with large counts can have a strong effect on the squared-error criterion. The selection-correction term may further increase this effect by generating large fitted values for some observations. This problem is especially important for the structural intercept, because estimation error in the second factorial moment is additionally transmitted through \(\widehat{\sigma}^2\).

\begin{figure}[H]
    \centering
    \includegraphics[width=0.72\textwidth]{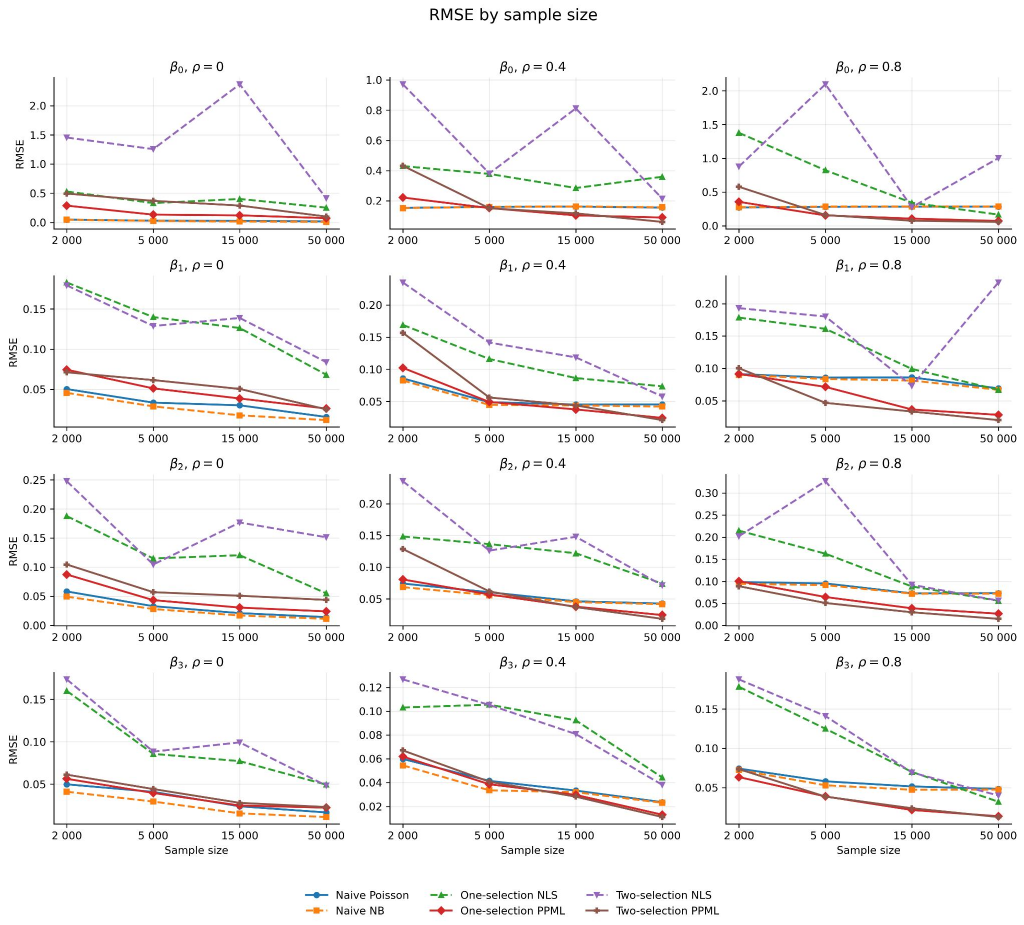}
    \caption{RMSE of the Monte Carlo estimators by sample size and dependence level}
    \label{fig:mc_rmse}
\end{figure}

Figure 3 shows that the median performance of the NLS estimators is much better than their RMSE performance. For example, at \(\rho=0.8\) the RMedSE of the two-selection NLS intercept decreases from approximately \(0.15\) to \(0.05\) as \(n\) increases, while its RMSE remains very large in some scenarios. The large difference between RMSE and RMedSE means that the NLS results are driven by a limited number of replications with extremely large estimation errors. This pattern is consistent with the instability discussed above: because NLS minimizes squared errors, a small number of observations with large counts or large fitted values may have a strong effect on the estimates. Thus, the poor RMSE performance of NLS appears to be driven mainly by rare but very large estimation errors rather than by systematically poor performance across replications.

The PPML estimators do not show the same degree of instability. Their RMSE and RMedSE decline more regularly with the sample size. At $\rho=0.8$, the RMedSE of the two-selection PPML intercept falls from approximately $0.07$ at $n=5{,}000$ to $0.02$ at $n=50{,}000$. In contrast, the RMedSE of the naive intercept remains close to $0.27$. Thus, the improvement of the PPML estimator is present not only in the mean squared error but also in the typical Monte Carlo replication.

\begin{figure}[H]
    \centering
    \includegraphics[width=0.72\textwidth]{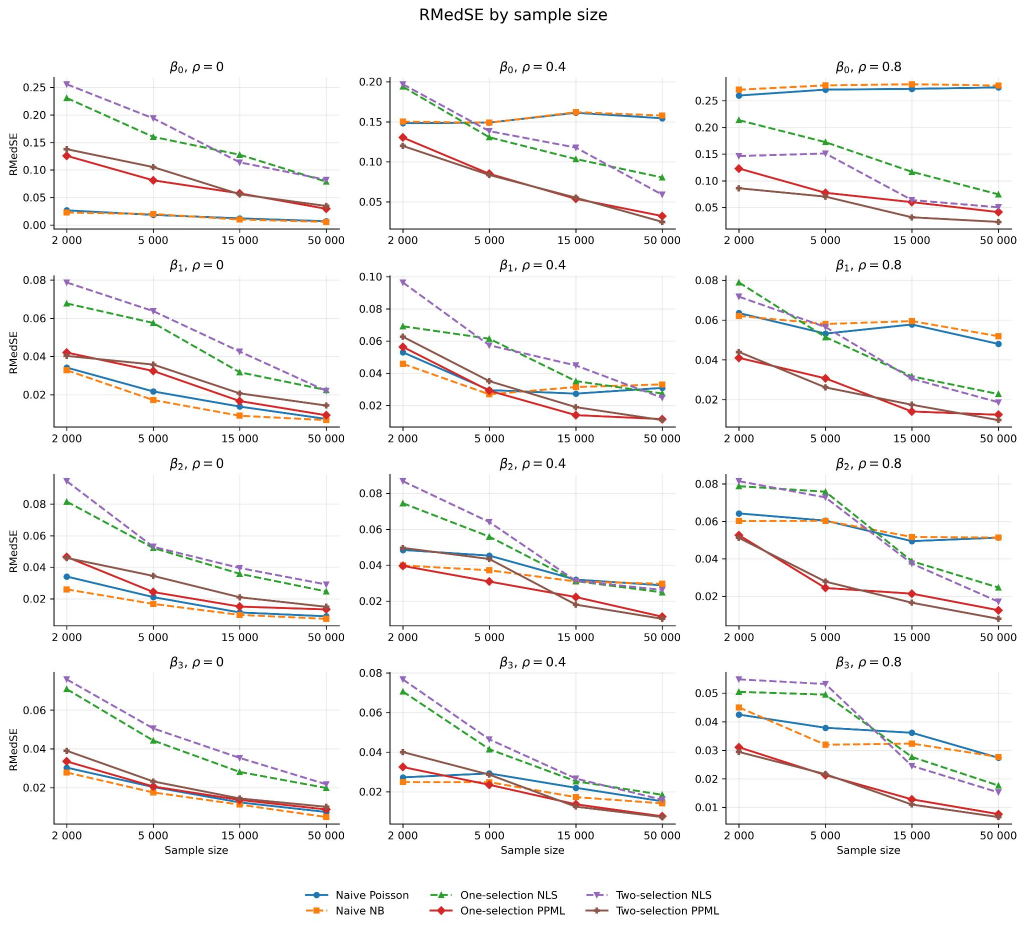}
    \caption{RMedSE of the Monte Carlo estimators by sample size and dependence level}
    \label{fig:mc_rmedse}
\end{figure}

Figure 4 confirms these conclusions using the share of replications in which each estimator has the smallest absolute error. When $\rho=0$, the naive negative binomial estimator has the highest win rate for most coefficients. As $\rho$ increases, its win rate falls, while the win rate of the PPML selection estimators rises.

For $\rho=0.4$, the win rate of the two-selection PPML estimator generally increases with $n$. For example, its win rate for $\beta_1$ increases from approximately $12\%$ at $n=5{,}000$ to $37\%$ at $n=50{,}000$. For $\rho=0.8$, the two-selection PPML estimator becomes the most frequent winner for most coefficients. Its win rate reaches approximately $45\%$ for the intercept, $43\%$ for $\beta_2$, and $37\%$ for $\beta_3$ at $n=50{,}000$.

Overall, the simulations show a clear trade-off. When selection dependence is absent, the naive estimators are more efficient. When the dependence is moderate or strong, their bias does not disappear with the sample size. In these cases, the two-selection PPML estimator becomes increasingly preferable as $n$ grows. The NLS version corrects part of the selection bias in typical replications, but its sensitivity to a small number of extreme estimates makes it less reliable than PPML.

\begin{figure}[H]
    \centering
    \includegraphics[width=0.72\textwidth]{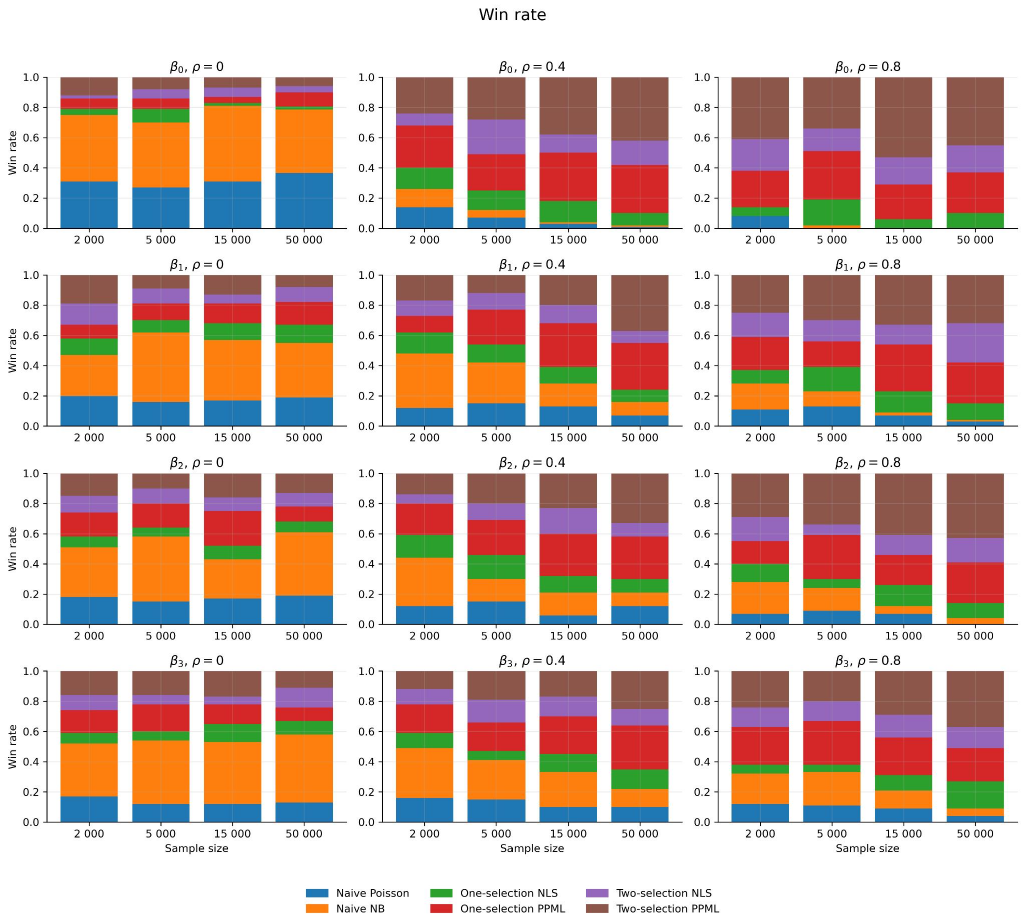}
    \caption{Win rate of the Monte Carlo estimators by sample size and dependence level}
    \label{fig:mc_winrate}
\end{figure}

\section{Conclusion}

This paper proposes a generalization of the count data sample selection model to the case of Poisson regression and multiple selection equations. The Monte Carlo results show that the proposed estimator outperforms both the naive count-data models and the standard single-selection Poisson model when the outcome is affected by more than one non-random selection. When the outcome and selection errors are not correlated, the naive estimators remain more efficient because no selection correction is required. However, when this dependence is moderate or strong, the bias of the naive estimators does not disappear as the sample size increases, while the proposed estimator provides substantially more accurate estimates.

The comparison of the two estimation methods used in the second step, NLS and PPML, gives a clear advantage to PPML. The accuracy of nonlinear least squares improves as the sample size grows, but NLS requires considerably more observations to obtain results comparable to PPML. It is also more sensitive to individual replications with very large estimation errors. In contrast, PPML provides stable and accurate estimates already in moderate samples. Therefore, PPML is the preferred second-step procedure in the considered simulation designs.

Moreover, the proposed approach is not limited to the Poisson distribution. For another discrete outcome distribution, the main estimation procedure remains unchanged, while only the moment condition used to recover the variance and covariance components must be adjusted.

The proposed method also provides information that is not available from naive count-data models. In addition to the regression coefficients, it allows the researcher to estimate the variances, covariances, and correlations between the errors of the selection equations and the error of the outcome equation. The estimated correlations show whether unobserved factors that increase the probability of selection also increase or decrease the expected outcome. Therefore, the proposed model allows the researcher to measure both the direction and the strength of selection on unobservables. Overall, the PPML-based estimator corrects the regression coefficients for multiple non-random selection processes and, at the same time, measures the dependence between these processes and the outcome equation. This makes it a useful tool for empirical applications in which the outcome is observed only after several selection decisions and gives researchers more tools for economic processes modeling. 

A natural extension of the framework is to consider endogenous switching models. Terza (1998) studies count-data models with a binary switch generated by a single latent equation, while the multiple-selection structure developed here can be extended to settings in which switching depends on several correlated latent processes. Future research may also consider PPML estimation in such models, including PPML for regime-specific conditional means.

\bibliographystyle{apalike}
\bibliography{references}

\pagebreak{}

\appendix

\section{Moment-generating function used for the selection correction}
\label{app:mgf}

For a zero-mean multivariate normal vector \(X\sim \mathcal{N}(0,\Sigma_{n\times n})\), truncated on the rectangular region \(a\leq X\leq b\), denote \(\alpha=\P\{a\leq X\leq b\}\). Following \citep{wilhelm2021} formula 5, the moment-generating function of the truncated vector is
\begin{equation}
m(t)
=
\E[\exp\{t'X\}\mid a\leq X\leq b]
=
e^W\Phi_{\alpha\Sigma}.
\label{eq:truncated_mgf}
\end{equation}
where \(W=\frac{1}{2}t'\Sigma t\) and
\begin{equation}
\Phi_{\alpha\Sigma}
=
\frac{1}{\alpha(2\pi)^{d/2}|\Sigma|^{1/2}}
\int_{a-\Sigma t}^{b-\Sigma t}
\exp\left\{
-\frac{1}{2}x'\Sigma^{-1}x
\right\}dx.
\end{equation}

\section{Identification without an exclusion restriction}
\label{app:identification}

This appendix provides a formal proof of the identification result stated in Section~\ref{sec:identification}. Throughout the proof, A1--A4 are assumed to hold.

Let $Z$ be the $k$-dimensional random vector of continuous regressors and let $z\in\mathbb{R}^k$ denote its realization. Define

$$
x(z)=
\begin{pmatrix}
1\\
z
\end{pmatrix},
\qquad
\eta(z)=c+Az.
$$

After absorbing the common factor $\exp\{\dfrac{\sigma^2}{2}\}$ into the intercept of the outcome equation, the conditional mean for selected observations is

$$
\mu(z;\beta,\delta)
=
\exp\{x(z)'\beta\}
\frac{\Phi_m(c+Az+\delta;R)}
{\Phi_m(c+Az;R)}.
$$

Suppose that two parameter pairs $(\beta,\delta)$ and
$(\widetilde{\beta},\widetilde{\delta})$ generate the same conditional mean for every $z$ in the support of $Z$. Then

$$
x(z)'(\beta-\widetilde{\beta})
=
\ln\Phi_m(c+Az+\widetilde{\delta};R)
-
\ln\Phi_m(c+Az+\delta;R).
$$

Define

$$
F(z)
=
x(z)'(\beta-\widetilde{\beta})
-
\ln\Phi_m(c+Az+\widetilde{\delta};R)
+
\ln\Phi_m(c+Az+\delta;R).
$$

By A2--A3, $F(\cdot)$ is real analytic and equals zero on a nonempty open set. Hence, by the identity theorem for real-analytic functions,

$$
F(z)=0
\qquad
\text{for every }z\in\mathbb{R}^k.
$$

Partition the outcome coefficient vectors as

$$
\beta=
\begin{pmatrix}
\beta_0\\
\beta_z
\end{pmatrix},
\qquad
\widetilde{\beta}=
\begin{pmatrix}
\widetilde{\beta}_0\\
\widetilde{\beta}_z
\end{pmatrix},
$$

where $\beta_0$ and $\widetilde{\beta}_0$ are the intercepts and
$\beta_z$ and $\widetilde{\beta}_z$ are the coefficient vectors associated with $z$. Let

$$
\Delta\beta_0=\beta_0-\widetilde{\beta}_0,
\qquad
\Delta\beta_z=\beta_z-\widetilde{\beta}_z.
$$

By A4, there exists $v\in\mathbb{R}^k$ such that

$$
Av=1_m.
$$

For any $h\in\mathbb{R}^k$, define

$$
z(t)=h+tv,
\qquad
t\in\mathbb{R}.
$$

Then

$$
c+Az(t)=c+Ah+t1_m.
$$

Thus, as $t\to+\infty$, every component of the selection index tends to $+\infty$, and

$$
\Phi_m(c+Ah+t1_m+\delta;R)\to1,
\qquad
\Phi_m(c+Ah+t1_m+\widetilde{\delta};R)\to1.
$$

Since $F(z(t))=0$ for every $t$,

$$
\begin{aligned}
0
&=
\lim_{t\to+\infty} F(z(t)) \\
&=
\lim_{t\to+\infty}
\Big[
\Delta\beta_0+h'\Delta\beta_z+t\,v'\Delta\beta_z \\
&-\ln\Phi_m(c+Ah+t1_m+\widetilde{\delta};R)
+\ln\Phi_m(c+Ah+t1_m+\delta;R)
\Big].
\end{aligned}
$$

Since
$$
\lim_{t\to+\infty}\ln\Phi_m(c+Ah+t1_m+\widetilde{\delta};R)=0,
\qquad
\lim_{t\to+\infty}\ln\Phi_m(c+Ah+t1_m+\delta;R)=0,
$$

the limit can be finite only if

$$
v'\Delta\beta_z=0.
$$

Therefore,

$$
\lim_{t\to+\infty}F(z(t))
=
\Delta\beta_0+h'\Delta\beta_z = 0
\qquad
\text{for every }h\in\mathbb{R}^k.
$$

The two logarithmic terms converge to zero as $t\to+\infty$. Therefore,

$$
v'\Delta\beta_z=0,
$$

and the same limit gives

$$
\Delta\beta_0+h'\Delta\beta_z=0
\qquad
\text{for every }h\in\mathbb{R}^k.
$$

Taking $h=0$ gives $\Delta\beta_0=0$, while taking $h=e_\ell$ for
$\ell=1,\ldots,k$ gives $\Delta\beta_{z,\ell}=0$. Hence

$$
\widetilde{\beta}=\beta.
$$

It remains to identify $\delta$. The equality of $\widetilde{\beta}=\beta$ now implies

$$
\Phi_m(c+Az+\widetilde{\delta};R)
=
\Phi_m(c+Az+\delta;R)
\qquad
\text{for every }z\in\mathbb{R}^k.
$$

By A4, $\eta(z)=c+Az$ can take any value in $\mathbb{R}^m$. Therefore,

$$
\Phi_m(\eta(z)+\widetilde{\delta};R)
=
\Phi_m(\eta(z)+\delta;R)
$$

for every $\eta(z)\in\mathbb{R}^m$.

Fix any $j\in{1,\ldots,m}$ and any $r\in\mathbb{R}$. Set

$$
\eta_j(z)=r
$$

and let

$$
\eta_\ell(z)\to+\infty
\qquad
\text{for every }\ell\neq j.
$$

Then

$$
\Phi(r+\widetilde{\delta}_j)
=
\Phi(r+\delta_j).
$$

Since the standard normal CDF is strictly increasing,

$$
\widetilde{\delta}_j=\delta_j.
$$

This holds for every $j$, and therefore

$$
\widetilde{\delta}=\delta.
$$

Hence,

$$
\widetilde{\beta}=\beta,
\qquad
\widetilde{\delta}=\delta,
$$

which proves the identification result.

\section{Covariance matrix derivation}

Let
\[
m_i(\beta,\rho,\gamma,R)
=
\exp\{x_i\beta+0.5\}\lambda(w_i\gamma,\rho;R)
\]
be the conditional mean of the outcome equation on the selected sample. Let $I_i=1\{z_{i,1}=\cdots=z_{i,m}=1\}$. The first step estimates $\gamma$ and $R$ from the likelihood function of the multivariate probit selection system. The corresponding first-order condition is
$
\sum_{i=1}^{n}
\dfrac{\partial \ell_i(\gamma,R)}{\partial(\gamma,R)}
=
0,
$
where the derivative with respect to $R$ is understood as the derivative with respect to the free correlation parameters of $R$.

The second step estimates $\beta$ and $\rho$ by nonlinear least squares. Its first-order condition is
$
\sum_{i=1}^{n}
\mathbb{I}\{Z_i=1\}
\dfrac{\partial m_i(\beta,\rho,\gamma,R)}{\partial(\beta,\rho)}
\left[
y_i-m_i(\beta,\rho,\gamma,R)
\right]
=
0.
$

Collect these two conditions into one vector:
\[
\psi_i(\theta)
=
\left(
\begin{array}{c}
\displaystyle
\frac{\partial \ell_i(\gamma,R)}{\partial(\gamma,R)}
\\[8pt]
\displaystyle
\mathbb{I}\{Z_i=1\}
\frac{\partial m_i(\beta,\rho,\gamma,R)}{\partial(\beta,\rho)}
\left[
y_i-m_i(\beta,\rho,\gamma,R)
\right]
\end{array}
\right),
\]
where $\theta$ collects $\beta$, $\gamma$, $\rho$, and the free correlation parameters of $R$. At the true value $\theta_0$,
$
E[\psi_i(\theta_0)]=0.
$
The estimator satisfies the sample analogue
$
\dfrac{1}{n}\sum_{i=1}^{n}\psi_i(\widehat{\theta})=0.
$

To obtain the asymptotic covariance matrix, expand the sample moment around $\theta_0$:
\[
\frac{1}{n}\sum_{i=1}^{n}\psi_i(\theta_0)
+
\left[
\frac{1}{n}\sum_{i=1}^{n}
\frac{\partial \psi_i(\theta_0)}{\partial\theta'}
\right]
(\widehat{\theta}-\theta_0)
+
o_p(n^{-1/2}) = 0.
\]
Therefore,
\[
\sqrt{n}(\widehat{\theta}-\theta_0)
=
-
H^{-1}
\dfrac{1}{\sqrt{n}}\sum_{i=1}^{n}\psi_i(\theta_0)
+
o_p(1),
\]
where
$
H
=
\E\left[
\dfrac{\partial \psi_i(\theta_0)}{\partial\theta'}
\right].
$
By the central limit theorem,
\[
\frac{1}{\sqrt{n}}\sum_{i=1}^{n}\psi_i(\theta_0)
\stackrel{d}{\to}
N(0,\Omega),
\]
where
$
\Omega
=
\E\left[
\psi_i(\theta_0)\psi_i(\theta_0)'
\right].
$
Hence,
\[
\sqrt{n}(\widehat{\theta}-\theta_0)
\stackrel{d}{\to}
\mathcal{N}\left(0,H^{-1}\Omega(H^{-1})'\right).
\]

Thus, the robust covariance matrix is estimated by
\[
\widehat{\mathrm{Var}}(\widehat{\theta})
=
\frac{1}{n}
\widehat{H}^{-1}
\widehat{\Omega}
(\widehat{H}^{-1})',
\]
where
\[
\widehat{H}
=
\frac{1}{n}
\sum_{i=1}^{n}
\frac{\partial \psi_i(\widehat{\theta})}{\partial\theta'},
\qquad
\widehat{\Omega}
=
\frac{1}{n}
\sum_{i=1}^{n}
\psi_i(\widehat{\theta})\psi_i(\widehat{\theta})'.
\]

This covariance matrix accounts for the generated correction term because the second-step condition depends on $\widehat{\gamma}$ and $\widehat{R}$ through $\lambda(w_i\widehat{\gamma},\rho;\widehat{R})$.

\section{Derivation of the decomposition formulas}

\label{app:decomp}

Denote the multiple selection event by \(Z_i=1\), where \(Z_i=1\) if and only if \(z_{i,1}=\cdots=z_{i,m}=1\). For selected observations, \(y_i=y_i^*\). Let \((y)_r=y(y-1)\cdots(y-r+1)\) denote the \(r\)-th falling factorial.

The factorial moment formula is applied before integrating out the latent outcome error. Conditional on \(x_i\) and \(\varepsilon_i\), the latent count outcome follows a Poisson distribution with conditional mean \(\exp\{x_i'\beta+\varepsilon_i\}\). Therefore, its \(r\)-th conditional factorial moment is
\[
\mathbb{E}\left[(y_i^*)_r\mid x_i,\varepsilon_i\right]
=
\exp\left\{r\left(x_i'\beta+\varepsilon_i\right)\right\}.
\]

The selected-sample factorial moment is obtained by applying the law of iterated expectations over \(\varepsilon_i\). Since \(y_i^*\) is conditionally independent of the selection errors given \(x_i\) and \(\varepsilon_i\), we have
\begin{align}
F_{r,i}
=
\mathbb{E}\left[(y_i^*)_r\mid x_i,w_i,Z_i=1\right]
\nonumber
=
\mathbb{E}\left[
\mathbb{E}\left[(y_i^*)_r\mid x_i,\varepsilon_i\right]
\Bigm|
x_i,w_i,Z_i=1
\right].
\end{align}

It follows that
\[
F_{r,i}
=
\exp\left\{r x_i'\beta\right\}
\mathbb{E}\left[
\exp\left\{r\varepsilon_i\right\}
\mid x_i,w_i,Z_i=1
\right].
\]

The event \(Z_i=1\) is equivalent to \(u_i\geq-w_i'\gamma\), where the inequality is understood componentwise. Since the outcome error and the selection errors are jointly normal,
\[
\mathbb{E}\left[
\exp\left\{r\varepsilon_i\right\}
\mid x_i,w_i,Z_i=1
\right]
=
\exp\left\{\dfrac{r^2\sigma^2}{2}\right\}
\dfrac{
\Phi_m\left(w_i'\gamma+r\delta;R\right)
}{
\Phi_m\left(w_i'\gamma;R\right)
}.
\]

Define the selection correction of order \(r\) as
\[
\lambda_{r,i}
=
\dfrac{
\Phi_m\left(w_i'\gamma+r\delta;R\right)
}{
\Phi_m\left(w_i'\gamma;R\right)
}.
\]
Then
\[
F_{r,i}
=
\exp\left\{r x_i'\beta\right\}
\exp\left\{\dfrac{r^2\sigma^2}{2}\right\}
\lambda_{r,i}.
\]

For \(r=1\), the conditional mean is \(\mu_i=\mathbb{E}[y_i^*\mid x_i,w_i,Z_i=1]\), where
\[
\mu_i
=
\exp\left\{x_i'\beta\right\}
\exp\left\{\dfrac{\sigma^2}{2}\right\}
\lambda_{1,i}.
\]
Consequently,
\[
\mu_i^r
=
\exp\left\{r x_i'\beta\right\}
\exp\left\{\dfrac{r\sigma^2}{2}\right\}
\lambda_{1,i}^r.
\]

The next step removes from \(F_{r,i}\) the terms already contained in the conditional mean and the selection correction. The term \(\mu_i^r\) removes the regression component \(\exp\{r x_i'\beta\}\), while the ratio \(\lambda_{1,i}^r/\lambda_{r,i}\) removes the selection correction of order \(r\). Therefore, define
\begin{align}
A_{r,i}
&=
\mathbb{E}\left[
\dfrac{(y_i^*)_r}{\mu_i^r}
\dfrac{\lambda_{1,i}^r}{\lambda_{r,i}}
\Bigm|
x_i,w_i,Z_i=1
\right]
\nonumber\\
&=
\dfrac{F_{r,i}}{\mu_i^r}
\dfrac{\lambda_{1,i}^r}{\lambda_{r,i}}.
\end{align}

Substituting the expressions for \(F_{r,i}\) and \(\mu_i^r\), we obtain
\[
A_{r,i}
=
\exp\left\{
\dfrac{r^2-r}{2}\sigma^2
\right\}.
\]

Thus, \(A_{r,i}\) does not depend on \(x_i\), \(w_i\), \(\beta\), \(\gamma\), \(R\), or \(\delta\). It depends only on the variance of the latent outcome error.

For \(r=2\), we have \(A_{2,i}=\exp\{\sigma^2\}\). Therefore, the outcome-error variance is identified as
\[
\sigma^2=\ln A_{2,i}.
\]

Since \(\delta_j=\sigma\rho_j\), the correlation between the outcome error and the \(j\)-th selection error is
\[
\rho_j=\dfrac{\delta_j}{\sigma},
\qquad j=1,\ldots,m.
\]

In practice, the population quantity is replaced by its sample analogue:
\[
\widehat{A}_2
=
\dfrac{1}{n_s}
\sum_{i:Z_i=1}
\dfrac{(y_i)_2}{\widehat{\mu}_i^2}
\dfrac{
\widehat{\lambda}_{1,i}^2
}{
\widehat{\lambda}_{2,i}
},
\]
where \(n_s\) is the number of selected observations, \(\widehat{\mu}_i\) is the fitted conditional mean, and
\[
\widehat{\lambda}_{r,i}
=
\dfrac{
\Phi_m\left(
w_i'\widehat{\gamma}
+
r\widehat{\delta};
\widehat{R}
\right)
}{
\Phi_m\left(
w_i'\widehat{\gamma};
\widehat{R}
\right)
}.
\]

The estimator of the outcome-error variance is \(\widehat{\sigma}^2=\ln\widehat{A}_2\), and hence \(\widehat{\sigma}=\sqrt{\ln\widehat{A}_2}\). The estimated correlations are
\[
\widehat{\rho}_j
=
\dfrac{\widehat{\delta}_j}{\widehat{\sigma}},
\qquad j=1,\ldots,m.
\]

These sample quantities provide the decomposition formulas used in the main text. The same general idea may be extended to other count-data distributions by using the corresponding factorial moment formulas.

\subsection{Standard errors for the decomposed parameters}

The decomposed parameters are functions of the quantities estimated in the previous steps. Therefore, their standard errors can be obtained by the delta method. Let
\[
\widehat\psi
=
(\widehat\beta',\widehat\gamma',\widehat r_R',\widehat\delta',
\widehat A_2,\widehat A_3)'
\]
be the vector of plug-in estimates used in the decomposition, where \(r_R\) contains the free off-diagonal elements of the correlation matrix \(R\). Let \(\widehat\vartheta=h(\widehat\psi)\) denote the vector of recovered structural parameters. In the negative binomial case, \(\widehat\vartheta\) contains \(\widehat\kappa\), \(\widehat\sigma^2\), \(\widehat\sigma\), and \(\widehat\rho\). In the Poisson case, the same formula is used with \(\kappa=0\).

By the delta method,
\[
\widehat{\Var}(\widehat\vartheta)
=
\widehat G
\widehat{\Var}(\widehat\psi)
\widehat G',
\qquad
\widehat G
=
\dfrac{\partial h(\psi)}
{\partial \psi'}
\bigg|_{\psi=\widehat\psi}.
\]
The standard errors are the square roots of the diagonal elements of \(\widehat{\Var}(\widehat\vartheta)\).

The covariance matrix \(\widehat{\Var}(\widehat\psi)\) should account for the fact that \(\widehat A_2\) and \(\widehat A_3\) are computed using generated estimates \(\widehat\beta\), \(\widehat\gamma\), \(\widehat R\), and \(\widehat\delta\). Formally, this can be done by stacking the first-step selection equations, the second-step conditional-mean equations, and the sample moment equations for \(\widehat A_2\) and \(\widehat A_3\). Equivalently, one may use a non-parametric bootstrap. In each bootstrap sample, the selection equations, the conditional-mean equation, and the decomposition formulas are re-estimated. The standard errors are then computed as the sample standard deviations of the recovered structural parameters across bootstrap replications.

\section{Monte Carlo simulation results}

\label{app:mntc}

\begin{sidewaystable}[p]
\centering
\scriptsize
\setlength{\tabcolsep}{2.5pt}
\caption{Monte Carlo results for $\rho=0$}
\label{tab:mc_results_rho_0_0}
\resizebox{\textwidth}{!}{%
\begin{tabular}{rlrrrrrrrrrrrrrrrr}
\toprule
 & & \multicolumn{4}{c}{$\beta_0$} & \multicolumn{4}{c}{$\beta_1$} & \multicolumn{4}{c}{$\beta_2$} & \multicolumn{4}{c}{$\beta_3$} \\
\cmidrule(lr){3-6}\cmidrule(lr){7-10}\cmidrule(lr){11-14}\cmidrule(lr){15-18}
$n$ & Model & Bias & RMSE & RMedSE & Win rate (\%) & Bias & RMSE & RMedSE & Win rate (\%) & Bias & RMSE & RMedSE & Win rate (\%) & Bias & RMSE & RMedSE & Win rate (\%) \\
\midrule
2\,000 & Naive Poisson & -0.000 & 0.048 & 0.027 & 31.0 & 0.004 & 0.050 & 0.034 & 20.0 & 0.010 & 0.058 & 0.034 & 18.0 & -0.002 & 0.050 & 0.030 & 17.0 \\
2\,000 & Naive NB & -0.001 & 0.045 & 0.023 & 44.0 & 0.001 & 0.046 & 0.033 & 27.0 & 0.006 & 0.049 & 0.026 & 33.0 & 0.003 & 0.041 & 0.028 & 35.0 \\
2\,000 & One-selection NLS & -0.043 & 0.530 & 0.231 & 4.0 & 0.023 & 0.183 & 0.068 & 11.0 & 0.027 & 0.188 & 0.082 & 7.0 & 0.003 & 0.160 & 0.071 & 7.0 \\
2\,000 & One-selection PPML & 0.029 & 0.288 & 0.126 & 7.0 & 0.002 & 0.074 & 0.042 & 9.0 & 0.009 & 0.088 & 0.047 & 16.0 & 0.002 & 0.056 & 0.034 & 15.0 \\
2\,000 & Two-selection NLS & 0.231 & 1.454 & 0.256 & 2.0 & 0.025 & 0.179 & 0.079 & 14.0 & 0.040 & 0.248 & 0.094 & 11.0 & -0.008 & 0.174 & 0.076 & 10.0 \\
2\,000 & Two-selection PPML & 0.103 & 0.495 & 0.138 & 12.0 & 0.003 & 0.071 & 0.040 & 19.0 & 0.022 & 0.105 & 0.046 & 15.0 & 0.002 & 0.061 & 0.039 & 16.0 \\
\midrule
5\,000 & Naive Poisson & -0.001 & 0.030 & 0.018 & 27.0 & 0.000 & 0.034 & 0.022 & 16.0 & -0.002 & 0.033 & 0.021 & 15.0 & -0.000 & 0.041 & 0.020 & 12.0 \\
5\,000 & Naive NB & -0.003 & 0.027 & 0.020 & 43.0 & -0.000 & 0.029 & 0.017 & 46.0 & 0.000 & 0.028 & 0.017 & 43.0 & -0.000 & 0.030 & 0.018 & 42.0 \\
5\,000 & One-selection NLS & -0.018 & 0.331 & 0.160 & 9.0 & 0.022 & 0.140 & 0.058 & 8.0 & -0.021 & 0.115 & 0.052 & 6.0 & 0.011 & 0.086 & 0.044 & 6.0 \\
5\,000 & One-selection PPML & 0.006 & 0.133 & 0.081 & 7.0 & 0.008 & 0.051 & 0.032 & 11.0 & -0.004 & 0.043 & 0.024 & 16.0 & 0.003 & 0.040 & 0.021 & 18.0 \\
5\,000 & Two-selection NLS & 0.203 & 1.256 & 0.194 & 6.0 & 0.007 & 0.129 & 0.064 & 10.0 & -0.019 & 0.105 & 0.053 & 10.0 & 0.007 & 0.088 & 0.051 & 6.0 \\
5\,000 & Two-selection PPML & 0.054 & 0.370 & 0.105 & 8.0 & 0.005 & 0.062 & 0.036 & 9.0 & -0.008 & 0.057 & 0.035 & 10.0 & 0.004 & 0.044 & 0.023 & 16.0 \\
\midrule
15\,000 & Naive Poisson & -0.000 & 0.027 & 0.012 & 31.0 & -0.003 & 0.030 & 0.014 & 17.0 & 0.001 & 0.021 & 0.011 & 17.0 & -0.005 & 0.024 & 0.012 & 12.0 \\
15\,000 & Naive NB & -0.002 & 0.016 & 0.010 & 50.0 & 0.001 & 0.018 & 0.009 & 40.0 & 0.002 & 0.017 & 0.010 & 26.0 & -0.003 & 0.016 & 0.011 & 41.0 \\
15\,000 & One-selection NLS & 0.016 & 0.402 & 0.128 & 2.0 & -0.008 & 0.126 & 0.032 & 11.0 & 0.003 & 0.121 & 0.036 & 9.0 & -0.008 & 0.077 & 0.028 & 12.0 \\
15\,000 & One-selection PPML & -0.011 & 0.121 & 0.057 & 4.0 & -0.008 & 0.039 & 0.017 & 13.0 & 0.002 & 0.031 & 0.015 & 23.0 & -0.003 & 0.025 & 0.014 & 13.0 \\
15\,000 & Two-selection NLS & 0.397 & 2.368 & 0.114 & 6.0 & -0.004 & 0.139 & 0.043 & 6.0 & 0.004 & 0.177 & 0.040 & 9.0 & -0.013 & 0.099 & 0.035 & 5.0 \\
15\,000 & Two-selection PPML & 0.045 & 0.289 & 0.056 & 7.0 & -0.005 & 0.051 & 0.021 & 13.0 & 0.004 & 0.051 & 0.021 & 16.0 & -0.003 & 0.028 & 0.014 & 17.0 \\
\midrule
50\,000 & Naive Poisson & 0.002 & 0.018 & 0.007 & 36.5 & 0.000 & 0.016 & 0.007 & 19.0 & 0.002 & 0.014 & 0.009 & 19.0 & 0.001 & 0.017 & 0.007 & 13.0 \\
50\,000 & Naive NB & 0.000 & 0.009 & 0.006 & 42.0 & 0.000 & 0.012 & 0.007 & 36.0 & 0.001 & 0.011 & 0.007 & 42.0 & -0.000 & 0.011 & 0.005 & 45.0 \\
50\,000 & One-selection NLS & 0.037 & 0.254 & 0.079 & 2.0 & 0.003 & 0.068 & 0.022 & 12.0 & 0.007 & 0.055 & 0.025 & 7.0 & -0.000 & 0.049 & 0.020 & 9.0 \\
50\,000 & One-selection PPML & 0.008 & 0.074 & 0.029 & 9.5 & 0.000 & 0.026 & 0.009 & 15.0 & 0.003 & 0.024 & 0.013 & 10.0 & 0.000 & 0.022 & 0.009 & 9.0 \\
50\,000 & Two-selection NLS & 0.060 & 0.413 & 0.082 & 4.0 & -0.003 & 0.084 & 0.022 & 10.0 & 0.021 & 0.151 & 0.029 & 9.0 & -0.003 & 0.048 & 0.022 & 13.0 \\
50\,000 & Two-selection PPML & 0.010 & 0.101 & 0.035 & 6.0 & 0.001 & 0.025 & 0.014 & 8.0 & 0.007 & 0.044 & 0.015 & 13.0 & -0.001 & 0.023 & 0.010 & 11.0 \\
\bottomrule
\end{tabular}%
}
\end{sidewaystable}

\begin{sidewaystable}[p]
\centering
\scriptsize
\setlength{\tabcolsep}{2.5pt}
\caption{Monte Carlo results for $\rho=0.4$}
\label{tab:mc_results_rho_0_4}
\resizebox{\textwidth}{!}{%
\begin{tabular}{rlrrrrrrrrrrrrrrrr}
\toprule
 & & \multicolumn{4}{c}{$\beta_0$} & \multicolumn{4}{c}{$\beta_1$} & \multicolumn{4}{c}{$\beta_2$} & \multicolumn{4}{c}{$\beta_3$} \\
\cmidrule(lr){3-6}\cmidrule(lr){7-10}\cmidrule(lr){11-14}\cmidrule(lr){15-18}
$n$ & Model & Bias & RMSE & RMedSE & Win rate (\%) & Bias & RMSE & RMedSE & Win rate (\%) & Bias & RMSE & RMedSE & Win rate (\%) & Bias & RMSE & RMedSE & Win rate (\%) \\
\midrule
2\,000 & Naive Poisson & 0.142 & 0.153 & 0.148 & 14.0 & 0.016 & 0.086 & 0.053 & 12.0 & -0.007 & 0.074 & 0.048 & 12.0 & -0.002 & 0.060 & 0.027 & 16.0 \\
2\,000 & Naive NB & 0.144 & 0.151 & 0.150 & 12.0 & 0.014 & 0.083 & 0.046 & 36.0 & -0.006 & 0.069 & 0.040 & 32.0 & -0.004 & 0.055 & 0.025 & 33.0 \\
2\,000 & One-selection NLS & -0.079 & 0.430 & 0.194 & 14.0 & 0.008 & 0.169 & 0.069 & 14.0 & -0.004 & 0.148 & 0.074 & 15.0 & 0.008 & 0.103 & 0.071 & 10.0 \\
2\,000 & One-selection PPML & -0.029 & 0.221 & 0.130 & 28.0 & 0.005 & 0.102 & 0.056 & 11.0 & -0.002 & 0.081 & 0.040 & 21.0 & -0.004 & 0.062 & 0.032 & 19.0 \\
2\,000 & Two-selection NLS & 0.228 & 0.972 & 0.197 & 8.0 & 0.035 & 0.235 & 0.096 & 10.0 & 0.002 & 0.236 & 0.087 & 6.0 & 0.007 & 0.127 & 0.077 & 10.0 \\
2\,000 & Two-selection PPML & 0.101 & 0.432 & 0.120 & 24.0 & 0.036 & 0.157 & 0.063 & 17.0 & 0.003 & 0.129 & 0.050 & 14.0 & -0.004 & 0.067 & 0.040 & 12.0 \\
\midrule
5\,000 & Naive Poisson & 0.152 & 0.159 & 0.149 & 7.0 & -0.004 & 0.049 & 0.030 & 15.0 & -0.005 & 0.061 & 0.045 & 15.0 & -0.001 & 0.041 & 0.029 & 15.0 \\
5\,000 & Naive NB & 0.154 & 0.160 & 0.149 & 5.0 & -0.001 & 0.045 & 0.027 & 27.0 & -0.006 & 0.056 & 0.037 & 15.0 & -0.001 & 0.034 & 0.025 & 26.0 \\
5\,000 & One-selection NLS & 0.007 & 0.378 & 0.131 & 13.0 & -0.025 & 0.116 & 0.061 & 12.0 & 0.001 & 0.136 & 0.056 & 16.0 & -0.009 & 0.105 & 0.041 & 6.0 \\
5\,000 & One-selection PPML & -0.007 & 0.152 & 0.085 & 24.0 & -0.010 & 0.049 & 0.029 & 23.0 & 0.003 & 0.057 & 0.031 & 23.0 & -0.003 & 0.039 & 0.023 & 19.0 \\
5\,000 & Two-selection NLS & 0.087 & 0.381 & 0.138 & 23.0 & -0.020 & 0.142 & 0.057 & 11.0 & 0.026 & 0.126 & 0.064 & 11.0 & -0.005 & 0.105 & 0.046 & 15.0 \\
5\,000 & Two-selection PPML & 0.020 & 0.150 & 0.084 & 28.0 & -0.003 & 0.056 & 0.035 & 12.0 & 0.015 & 0.062 & 0.043 & 20.0 & -0.003 & 0.041 & 0.029 & 19.0 \\
\midrule
15\,000 & Naive Poisson & 0.159 & 0.161 & 0.161 & 3.0 & -0.002 & 0.045 & 0.027 & 13.0 & 0.001 & 0.046 & 0.032 & 6.0 & 0.002 & 0.033 & 0.022 & 10.0 \\
15\,000 & Naive NB & 0.160 & 0.162 & 0.162 & 1.0 & -0.003 & 0.044 & 0.031 & 15.0 & -0.001 & 0.045 & 0.031 & 15.0 & 0.001 & 0.032 & 0.017 & 23.0 \\
15\,000 & One-selection NLS & 0.024 & 0.284 & 0.103 & 14.0 & -0.003 & 0.086 & 0.035 & 11.0 & -0.006 & 0.122 & 0.031 & 11.0 & 0.004 & 0.092 & 0.025 & 12.0 \\
15\,000 & One-selection PPML & -0.009 & 0.103 & 0.054 & 32.0 & 0.003 & 0.038 & 0.014 & 29.0 & -0.003 & 0.038 & 0.022 & 28.0 & -0.001 & 0.030 & 0.014 & 25.0 \\
15\,000 & Two-selection NLS & 0.143 & 0.812 & 0.118 & 12.0 & -0.008 & 0.118 & 0.045 & 12.0 & -0.032 & 0.148 & 0.031 & 17.0 & 0.005 & 0.081 & 0.027 & 13.0 \\
15\,000 & Two-selection PPML & 0.007 & 0.116 & 0.055 & 38.0 & 0.004 & 0.044 & 0.019 & 20.0 & -0.002 & 0.037 & 0.018 & 23.0 & -0.002 & 0.028 & 0.012 & 17.0 \\
\midrule
50\,000 & Naive Poisson & 0.152 & 0.155 & 0.154 & 1.0 & 0.003 & 0.046 & 0.031 & 7.0 & -0.004 & 0.043 & 0.029 & 12.0 & -0.002 & 0.024 & 0.015 & 10.0 \\
50\,000 & Naive NB & 0.154 & 0.157 & 0.158 & 1.0 & 0.003 & 0.042 & 0.033 & 9.0 & -0.005 & 0.042 & 0.030 & 9.0 & -0.002 & 0.023 & 0.014 & 12.0 \\
50\,000 & One-selection NLS & -0.011 & 0.359 & 0.080 & 8.0 & -0.003 & 0.074 & 0.027 & 8.0 & 0.001 & 0.074 & 0.025 & 9.0 & -0.002 & 0.044 & 0.018 & 13.0 \\
50\,000 & One-selection PPML & -0.017 & 0.089 & 0.032 & 32.0 & 0.001 & 0.024 & 0.012 & 31.0 & 0.001 & 0.025 & 0.011 & 28.0 & -0.001 & 0.013 & 0.008 & 29.0 \\
50\,000 & Two-selection NLS & 0.032 & 0.214 & 0.059 & 16.0 & 0.005 & 0.058 & 0.025 & 8.0 & 0.001 & 0.071 & 0.026 & 9.0 & 0.002 & 0.038 & 0.016 & 11.0 \\
50\,000 & Two-selection PPML & 0.006 & 0.060 & 0.025 & 42.0 & 0.003 & 0.021 & 0.011 & 37.0 & 0.000 & 0.019 & 0.010 & 33.0 & 0.001 & 0.011 & 0.007 & 25.0 \\
\bottomrule
\end{tabular}%
}
\end{sidewaystable}

\begin{sidewaystable}[p]
\centering
\scriptsize
\setlength{\tabcolsep}{2.5pt}
\caption{Monte Carlo results for $\rho=0.8$}
\label{tab:mc_results_rho_0_8}
\resizebox{\textwidth}{!}{%
\begin{tabular}{rlrrrrrrrrrrrrrrrr}
\toprule
 & & \multicolumn{4}{c}{$\beta_0$} & \multicolumn{4}{c}{$\beta_1$} & \multicolumn{4}{c}{$\beta_2$} & \multicolumn{4}{c}{$\beta_3$} \\
\cmidrule(lr){3-6}\cmidrule(lr){7-10}\cmidrule(lr){11-14}\cmidrule(lr){15-18}
$n$ & Model & Bias & RMSE & RMedSE & Win rate (\%) & Bias & RMSE & RMedSE & Win rate (\%) & Bias & RMSE & RMedSE & Win rate (\%) & Bias & RMSE & RMedSE & Win rate (\%) \\
\midrule
2\,000 & Naive Poisson & 0.261 & 0.271 & 0.260 & 8.0 & -0.003 & 0.091 & 0.064 & 11.0 & 0.012 & 0.098 & 0.064 & 7.0 & 0.016 & 0.074 & 0.043 & 12.0 \\
2\,000 & Naive NB & 0.267 & 0.275 & 0.270 & 0.0 & 0.001 & 0.090 & 0.062 & 17.0 & 0.009 & 0.095 & 0.060 & 21.0 & 0.017 & 0.072 & 0.045 & 20.0 \\
2\,000 & One-selection NLS & 0.178 & 1.377 & 0.213 & 6.0 & -0.007 & 0.179 & 0.079 & 9.0 & 0.046 & 0.215 & 0.079 & 12.0 & -0.001 & 0.178 & 0.050 & 6.0 \\
2\,000 & One-selection PPML & 0.040 & 0.354 & 0.123 & 24.0 & -0.007 & 0.091 & 0.041 & 22.0 & 0.025 & 0.101 & 0.053 & 15.0 & 0.004 & 0.063 & 0.031 & 25.0 \\
2\,000 & Two-selection NLS & 0.247 & 0.876 & 0.146 & 21.0 & -0.005 & 0.193 & 0.072 & 16.0 & -0.011 & 0.202 & 0.081 & 16.0 & -0.007 & 0.188 & 0.055 & 13.0 \\
2\,000 & Two-selection PPML & 0.120 & 0.577 & 0.086 & 41.0 & 0.000 & 0.100 & 0.044 & 25.0 & -0.000 & 0.089 & 0.051 & 29.0 & 0.006 & 0.073 & 0.029 & 24.0 \\
\midrule
5\,000 & Naive Poisson & 0.274 & 0.280 & 0.271 & 0.0 & -0.006 & 0.086 & 0.053 & 13.0 & 0.012 & 0.095 & 0.060 & 9.0 & -0.003 & 0.058 & 0.038 & 11.0 \\
5\,000 & Naive NB & 0.278 & 0.283 & 0.279 & 2.0 & -0.007 & 0.084 & 0.058 & 10.0 & 0.012 & 0.091 & 0.060 & 15.0 & -0.005 & 0.053 & 0.032 & 22.0 \\
5\,000 & One-selection NLS & 0.046 & 0.822 & 0.172 & 17.0 & 0.000 & 0.161 & 0.051 & 16.0 & 0.008 & 0.163 & 0.076 & 6.0 & 0.005 & 0.125 & 0.049 & 5.0 \\
5\,000 & One-selection PPML & -0.017 & 0.153 & 0.078 & 32.0 & 0.006 & 0.071 & 0.031 & 17.0 & -0.004 & 0.065 & 0.024 & 29.0 & 0.002 & 0.039 & 0.021 & 29.0 \\
5\,000 & Two-selection NLS & 0.359 & 2.097 & 0.151 & 15.0 & 0.021 & 0.180 & 0.056 & 14.0 & 0.042 & 0.327 & 0.073 & 7.0 & -0.006 & 0.141 & 0.053 & 13.0 \\
5\,000 & Two-selection PPML & 0.013 & 0.159 & 0.070 & 34.0 & 0.005 & 0.047 & 0.026 & 30.0 & 0.006 & 0.051 & 0.028 & 34.0 & -0.000 & 0.038 & 0.022 & 20.0 \\
\midrule
15\,000 & Naive Poisson & 0.276 & 0.282 & 0.272 & 0.0 & -0.011 & 0.086 & 0.058 & 7.0 & 0.011 & 0.073 & 0.049 & 7.0 & 0.004 & 0.051 & 0.036 & 9.0 \\
15\,000 & Naive NB & 0.282 & 0.287 & 0.281 & 0.0 & -0.012 & 0.081 & 0.059 & 2.0 & 0.010 & 0.072 & 0.052 & 5.0 & 0.004 & 0.047 & 0.032 & 12.0 \\
15\,000 & One-selection NLS & -0.028 & 0.341 & 0.117 & 6.0 & 0.014 & 0.099 & 0.032 & 14.0 & -0.003 & 0.088 & 0.039 & 14.0 & 0.010 & 0.070 & 0.028 & 10.0 \\
15\,000 & One-selection PPML & -0.034 & 0.106 & 0.060 & 23.0 & 0.007 & 0.037 & 0.014 & 31.0 & -0.003 & 0.039 & 0.021 & 20.0 & 0.002 & 0.021 & 0.013 & 25.0 \\
15\,000 & Two-selection NLS & 0.033 & 0.275 & 0.064 & 18.0 & 0.003 & 0.073 & 0.031 & 13.0 & 0.005 & 0.092 & 0.037 & 13.0 & 0.005 & 0.069 & 0.025 & 15.0 \\
15\,000 & Two-selection PPML & -0.002 & 0.074 & 0.032 & 53.0 & -0.001 & 0.033 & 0.017 & 33.0 & 0.003 & 0.030 & 0.017 & 41.0 & -0.000 & 0.024 & 0.011 & 29.0 \\
\midrule
50\,000 & Naive Poisson & 0.279 & 0.283 & 0.275 & 0.0 & 0.001 & 0.069 & 0.048 & 3.0 & -0.013 & 0.073 & 0.051 & 0.0 & -0.004 & 0.048 & 0.027 & 4.0 \\
50\,000 & Naive NB & 0.282 & 0.286 & 0.278 & 0.0 & 0.002 & 0.067 & 0.052 & 1.0 & -0.012 & 0.072 & 0.051 & 4.0 & -0.005 & 0.047 & 0.028 & 5.0 \\
50\,000 & One-selection NLS & -0.018 & 0.165 & 0.075 & 10.0 & -0.011 & 0.067 & 0.023 & 11.0 & -0.000 & 0.056 & 0.025 & 10.0 & 0.005 & 0.032 & 0.018 & 18.0 \\
50\,000 & One-selection PPML & -0.025 & 0.071 & 0.041 & 27.0 & -0.001 & 0.028 & 0.012 & 27.0 & -0.001 & 0.027 & 0.013 & 27.0 & -0.000 & 0.013 & 0.008 & 22.0 \\
50\,000 & Two-selection NLS & 0.155 & 1.002 & 0.050 & 18.0 & 0.022 & 0.233 & 0.019 & 26.0 & -0.002 & 0.056 & 0.017 & 16.0 & 0.002 & 0.040 & 0.015 & 14.0 \\
50\,000 & Two-selection PPML & 0.015 & 0.057 & 0.023 & 45.0 & 0.003 & 0.020 & 0.010 & 32.0 & 0.000 & 0.015 & 0.008 & 43.0 & -0.000 & 0.012 & 0.007 & 37.0 \\
\bottomrule
\end{tabular}%
}
\end{sidewaystable}

\end{document}